\documentstyle[prl,aps,multicol,epsf]{revtex}

\renewcommand{\narrowtext}{\begin{multicols}{2}
\global\columnwidth20.5pc} 
\renewcommand{\widetext}{\end{multicols}
\global\columnwidth42.5pc} \multicolsep = 8pt plus 4pt minus 3pt

\begin{document}
\title{Exciton Beats in GaAs Quantum Wells: Bosonic Representation and Collective
Effects}
\author{J. Fern\'andez-Rossier and C. Tejedor}
\address{Departamento de F\'{\i }sica Te\'{o}rica de la Materia Condensada, \\
Universidad Aut\'{o}noma de Madrid, Cantoblanco, 28049 Madrid, Spain}
\author{R. Merlin}
\address{Department of Physics, The University of Michigan, \\
Ann Arbor, MI 48109-1120}
\date{\today}
\maketitle

\begin{abstract}
We discuss light-heavy hole beats observed in transient optical experiments
in GaAs quantum wells in terms of a free-boson coherent state model. This
approach is compared with descriptions based on few-level representations.
Results lead to an interpretation of the beats as due to classical
electromagnetic interference. The boson picture correctly describes photon
excitation of extended states and accounts for experiments involving
coherent control of the exciton density and Rayleigh scattering beating.
\end{abstract}



\narrowtext

The optical properties of semiconductor quantum wells (QW) and, in
particular, the coherent dynamics of excitons following resonant excitation
with ultrafast laser pulses have attracted much attention in recent years 
\cite
{QB,PBvsQB,Wang,Deveaud,Chemla,Stoltzbook,Baumberg,Shah,Cecc,Amand1,recent1,recent2}%
. It is generally accepted that there is a transfer of coherence between the
optical field and the QW that disappears in a characteristic time $T_{2}$
(picoseconds for GaAs) after the laser is turned off. However, the questions
as to how the coherence is actually induced and that of the nature of the
coherent state of the solid are poorly understood. In this paper, we address
these points by re-examining the long-standing problem of the (classical vs.
quantum) nature of the ubiquitous beats associated with the light-hole (LX)
and heavy-hole (HX) excitons, which are observed in transient optical
experiments on QW \cite{QB,PBvsQB,Wang,Deveaud}. To this end, we consider
the coherent behavior of excitons using the simplest albeit non-trivial
model where they are treated as non-interacting bosons. Hence, our work
relates directly to studies for which nonlinear effects are not important 
\cite{Wang,Deveaud,Baumberg,Amand1,recent1,recent2} but it is not aimed at
explaining the four-wave-mixing (FWM) experiments that dominate the field 
\cite{QB}. Nevertheless, since nonlinear effects are, typically, weak
compared with harmonic contributions, the free-boson picture provides in all
cases the correct lowest-order wavefunction of the photoexcited solid. It
should be emphasized that our results apply only to excitons in {\em %
weakly-localized}{\bf \ }states.

The structure of this paper is as follows. First, we discuss the bosonic
representation of excitons in QW's, obtaining the exact collective state of
a QW driven by an arbitrary laser pulse and show that its properties {\it %
vis-a-vis }coherence are identical to those of coherent optical fields \cite
{Glauber}. From this, it follows that laser-induced coherence is a
collective property of the exciton field that is not owed by individual
excitons. Using the many-exciton wavefunction, we provide a quantitative
description of recent experiments where two laser pulses are used to
coherently control the HX density in a GaAs QW\cite{Baumberg,Amand1}. We
also analyze the case where a single pulse excites both the LX and HX states
and argue that the resulting beats \cite
{QB,PBvsQB,Wang,Deveaud,recent1,recent2} are not due to (single-exciton)
quantum interference, as advocated by few-level models, but to polarization
interference associated with the emission of phased arrays of classical
antennas. Finally, we consider Rayleigh scattering experiments \cite
{Wang,Deveaud,recent1,recent2} and show that the bosonic approach accounts
for the quadratic rise in the intensity at short times that is observed in
the experiments \cite{Deveaud}.

At small electric fields and near band-gap excitation, the quanta of the
induced polarization field, ${\bf P}$, are the excitons \cite{Hopfield}.
Since their number is proportional to the illuminated volume of the sample, $%
V$, it is clear that the quantum description of an excited QW (and, as we
shall see, that of the beats) becomes, in some sense, a {\em many-exciton}
problem. Our discussion concerns itself with extended states. The
atomic-like scheme \cite{Yajima} where excitons are represented by a
collection of distinguishable few-level systems is the correct
representation in the strongly localized regime as in the case of quantum
dots or excitons bound to islands \cite{QD}. It should be emphasized that,
in the latter picture, the optically-induced coherence relies on intra-level
quantum entanglement and, therefore, that it is a {\em single- }or, at most,
a{\em \ few-exciton} effect. For our approach to work, the areal density of
photogenerated excitons, $n$, must be sufficiently low so that the excited
state of the solid can be described as a system of non-interacting bosons 
\cite{SR}. Hence, our discussion applies to a GaAs QW excited with low
intensity pulses using photon energies in the vicinity of the LX\ and HX
absorption lines. Note that, under these conditions, the bosonic picture
follows directly from the semiconductor Bloch equations \cite{SR} and
BCS-like fermionic theories \cite{Comte}. We are aware that our approach
ignores all but a small fraction of the QW Hilbert space. However, the
experiments considered here \cite
{Wang,Deveaud,Baumberg,Amand1,recent1,recent2} are well described by models
that rely on the same restricted basis and, thus, we conclude that the
neglected sectors of the Hilbert space ({\it e.g.}, the electron-hole
continuum) play only a secondary role in many cases of interest.

For a perfect QW (the weak-disorder limit is discussed later), the
Hamiltonian describing free excitons coupled to a classical electromagnetic
field is \cite{Jorda}: 
\begin{equation}
\widehat{H}=\sum_{{\bf k},\alpha ,M}\hbar \omega _{{\bf k},\alpha }A _{{\bf k%
},\alpha ,M}^{\dagger }A_{{\bf k},\alpha ,M}-\int {\bf P} \cdot {\bf E}({\bf %
r},t)\,dV  \label{hamil}
\end{equation}
where ${\bf E}$ is the electric field. $A_{{\bf k},\alpha ,M}^{\dagger }$ is
a {\em boson} operator that creates a QW exciton with in-plane momentum $%
{\bf k}$ valence band index $\alpha $ and pseudo angular momentum $M$.
Relevant to our problem are the lowest-lying optically active ($M=\pm 1$)
heavy- ($\alpha =H$) and light-hole ($\alpha =L$) QW states. The
single-particle energy is $\hbar \omega _{{\bf k},\alpha }=E_{\alpha
}^{g}-\epsilon _{\alpha }+K_{E}\,$where $K_{E}=\hbar ^{2}k^{2}/2m_{\alpha }$
is the center-of-mass kinetic energy, {\em m}$_{\alpha }$ is the exciton
mass,{\small \thinspace }$E_{\alpha }^{g}$ is the relevant QW gap and $%
\epsilon _{\alpha }$ is the exciton binding energy. Consider normal
incidence, {\it i.e.}, the light couples only to the state at ${\bf k=0}$.
Using the fact that typical QW widths are considerably smaller than the
light wavelength, we write the interaction term as $-V\sum_{M}P_{M}E_{M}$ 
where 
\begin{equation}
P_{M}=\frac{1}{\sqrt{V}}\sum_{\alpha }G_{\alpha ,M}\left( A_{{\bf 0,}\alpha
,M}^{\dagger }+A_{{\bf 0},\alpha ,M}\right)  \label{dipole}
\end{equation}
$\,$and $E_{M}({\bf r}=0,t)$ are, respectively, the $M=\pm 1$ components of $%
{\bf P}$ and ${\bf E}$, and $G_{\alpha ,M}$ are constants proportional to
the dipole matrix element \cite{SR}. The Hamiltonian (\ref{hamil}) is
equivalent to that of a set of{\em \ independent} harmonic oscillators (the
exciton $H\,$and $L\,$modes at ${\bf k=0}$) driven by an external field. For
arbitrary driving force and initial state, this problem can be solved
exactly by applying a time-dependent Glauber transformation \cite
{Glauber,Galindo}. In particular, if the QW is initially in its ground state
and the external field is turned on at $t=-\infty $ , the exact state of the
exciton field at time $t$ is \cite{Galindo} 
\begin{equation}
|\Xi \rangle =e^{-i\widehat{H}_{0}t/\hbar }\prod_{\alpha {\bf ,}M}e^{-\left|
K_{\alpha ,M}\right| ^{2}/2}e^{iK_{\alpha ,M}A_{{\bf 0},\alpha ,M}^{\dagger
}}|0\rangle  \label{coh}
\end{equation}
where $\widehat{H}_{0}$ is the free exciton Hamiltonian and 
\begin{equation}
K_{\alpha ,M}(t)=\frac{\sqrt{V}G_{\alpha ,M}}{2\hbar }\int_{-\infty
}^{t}E_{M}(s)e^{i\omega _{{\bf 0},\alpha }s}ds.  \label{K}
\end{equation}

The wavefunction (\ref{coh}) is a product of states of individual modes that
is formally identical to the so-called (multimode) {\em coherent state }%
proposed by Glauber as the quantum counterpart to classical light \cite
{Glauber}. As for the photon case, exciton coherent states are fully
characterized by the complex function $K_{\alpha ,M}(t)$ which defines the
classical phase. Because the system is not nonlinear, the induced
polarization is exactly given by 
\begin{equation}
\langle \Xi |P_{M}|\Xi \rangle =\frac{2}{\sqrt{V}}\sum_{\alpha }G_{\alpha
,M}Re\left( ie^{i\omega _{{\bf 0},\alpha }t}K_{\alpha ,M}(t)\right) .
\label{polar}
\end{equation}
while the areal density of $\alpha $-excitons with momenta ${\bf k}$ and $M$
is $n_{{\bf k},\alpha ,M}=(N_{{\bf 0,}\alpha ,M}l/V)\delta _{{\bf k},{\bf 0}%
} $where 
\begin{equation}
N_{{\bf 0,}\alpha ,M}=\langle \Xi |A_{{\bf 0,}\alpha ,M}^{\dagger }A_{{\bf 0,%
}\alpha ,M}|\Xi \rangle =|K_{\alpha ,M}(t)|^{2}  \label{densi}
\end{equation}
and $l{\sl \,}$is the width of the well. Here, we note that the linear
susceptibility, as easily obtained from (\ref{polar}), is identical to that
of the analogous few-level model. These expressions apply strictly to
non-interacting excitons. Coupling with the environment and interactions
between excitons lead to {\em dephasing} in that the pure state (\ref{coh})
evolves into a statistical mixture of coherent states with random phases. It
is beyond the scope of this work to provide a microscopic account of these
interactions. Instead, we will rely on the exponential decay approximation
and treat dephasing phenomenologically.

{\em Coherent control theory.--- } We now analyze recent experiments where
light pulses are used to control the exciton density in a GaAs-QW \cite
{Baumberg,Amand1}. Within the bosonic description, these results constitute
a striking demonstration of collective behavior. The experiments rely on two
phase-locked pulses tuned close to an exciton mode of energy $\hbar \omega $
and separated by a time delay $\tau \,$which serves as the control parameter
for the exciton density ($n=\sum_{{\bf k},\alpha ,M}n_{{\bf k,\alpha ,}M}$
is probed indirectly by monitoring the reflectivity of a third pulse \cite
{Baumberg} or the luminescence intensity \cite{Amand1}). The data can be
fitted to $n=2n_{s}[1+\cos (\omega \tau )e^{-\tau /T_{2}}]$ where $n_{s}$
represents the exciton density generated by a single pulse. Hence, small
changes in the time delay ($\pi /\omega $ $\approx 1$ fs) lead to large
variations of $n$ from zero, corresponding to destructive interference
between the pulses, to nearly four times the value for one pulse \cite
{Baumberg,Amand1}. These results can be easily explained using the
expressions derived previously. To account for the double pulse, we write $%
E=E_{1}(t)+E_{2}(t)$ where $E_{1}(t)=F(t)$, $E_{2}(t)=F(t-\tau )$ and 
\begin{equation}
F(t)=E_{0}\sin (\Omega t)e^{-(t/T)^{2}}.  \label{elec}
\end{equation}
$T$ and $\Omega \,$are the pulse width and central frequency. From (\ref{K}%
),(\ref{densi}) and (\ref{elec}), and assuming that the pulses couple only
to a single ${\bf k}={\bf 0}\,$mode of frequency $\omega \,$(the extension
to many modes is straightforward) we obtain the areal density 
\begin{equation}
n\simeq |K(\infty )|^{2}(l/V)=2n_{s}[1+\cos (\omega \tau )].  \label{control}
\end{equation}
$n_{s}=(gE{_{0}}T/\hbar )^{2}\exp -[\omega ^{2}T^{2}(1+r^{2})/4]\sinh
[r\omega ^{2}T^{2}/2]$ is the average density created by one pulse, ${\em g}$
is a constant and $r=\Omega /\omega $ measures the detuning between the
laser and the exciton resonance. The result (\ref{control}) contains the
essential feature of coherent control, namely, the oscillatory term. Decay
can be incorporated into the model by considering, e.g, a distribution of
modes of different energies \cite{JOCM}. It should be emphasized that, in
the coherent-state representation, control of the density follows from the
fact that the wavefunction (\ref{coh}) is a linear superposition of states
with different number of excitons (not surprisingly, the same result can be
obtained from a classical analysis of a kicked harmonic oscillator as it is
done in phonon control studies \cite{phonon}). Also note that the state
induced by the pair of pulses

\begin{equation}
|\Xi _{E_{1}+E_{2}}\rangle \sim e^{i(K_{1}+K_{2})A^{\dagger }}|0\rangle
\label{e1+e2}
\end{equation}
cannot be approximated by the superposition state $[1+i(K_{1}+K_{2})A^{%
\dagger }]|0\rangle $ because $K_{1,2}\propto \sqrt{V}$. Hence, our
interpretation differs significantly from that of the two-level model \cite
{Baumberg} where control is understood as a single-exciton quantum
interference effect.

{\em LX-HX beats.--- }We now turn to the main subject of this work, namely,
the beats of frequency $\omega _{L}-\omega _{H}$ ($\omega _{L}$\thinspace
and $\omega _{H}\,$are, respectively, the frequencies of the LX and HX
excitons) as reported in a wide variety of experiments \cite
{QB,PBvsQB,Wang,Deveaud,Shah}, which are conventionally characterized \cite
{QB,PBvsQB} as a quantum interference phenomenon much like the so-called
quantum beats of atomic physics \cite{Yajima}. Within the atomic-like
interpretation, the QW is treated as a set of three-level systems whose
excited states are the LX- and HX- states. The role of the optical pulses is
to bring each system into a {\em sum }state,{\em \ i.e.},{\em \ }a linear
combination of LX, HX and the ground state and, thus, the beats are a
consequence of intra-exciton quantum entanglement \cite{Yajima}. As we shall
see, the actual state of the solid is very different from that of the
atomic-like picture.

Consider first a perfect QW and circularly polarized pulses of bandwidth
large enough so that both LX- and HX-modes of well-defined angular momentum $%
M$ are resonantly excited. From (\ref{coh}), it directly follows that the
wavefunction in this case is a {\em product} of LX and HX coherent states 
\begin{equation}
|\Xi _{LH}\rangle \sim e^{-i\omega _{H}tN_{H}}e^{iK_{H}(t)A_{H}^{\dagger
}}e^{-i\omega _{L}tN_{L}}e^{iK_{L}(t)A_{L}^{\dagger }}|0\rangle  \label{LH}
\end{equation}
where $N_{\alpha }\equiv A_{\alpha }^{\dagger }A_{\alpha }$. Using (\ref
{polar}), it can be shown that (\ref{LH}) leads to LX-HX beats that reflect
interference of coherent light emitted by two phased arrays of antennas. We
note that the beating field, associated with $\langle \Xi _{LH}|P|\Xi
_{LH}\rangle \neq 0$, is classical in nature as opposed to the field due to
spontaneous emission that characterizes quantum beats \cite{Yajima}.
Moreover, because $\Xi _{LH}\,$cannot be expressed in terms of sums of LX
and HX states (since $K_{L,H}\propto \sqrt{V}$,{\it \ }this is not even
possible as an approximation), it is apparent that quantum-superposition
arguments cannot be used to describe beats associated with extended excitons.

{\em Rayleigh scattering.---} To consider the experiments on resonant
secondary emission involving elastic (Rayleigh) scattering \cite
{Wang,Deveaud,recent1,recent2}, we add to (\ref{hamil}) the term 
\begin{equation}
\widehat{U}=\sum_{\alpha ,{\bf k}\neq {\bf 0}}V_{\alpha }({\bf k})\left( A_{%
{\bf k,}\alpha }^{\dagger }A_{{\bf 0,}\alpha }+A_{{\bf k,}\alpha }A_{{\bf 0}%
,\alpha }^{\dagger }\right)   \label{scatt}
\end{equation}
describing the {\em elastic} scattering from the state coupled to the laser (%
${\bf k}={\bf 0}$) to other states (${\bf k}\neq 0$) due to interaction with
defects such as impurities and interface roughness. Since the disorder
described by (\ref{scatt}) does not affect the internal degrees of freedom,
our results do not apply to quantum dots \cite{QD}. $\widehat{U}$ gives rise
to Rayleigh scattering, {\it i.e.}, emission of photons with the same energy
but different in-plane momentum than the incident light \cite{teoray}.
Following \cite{Van}, we adopt the Heisenberg picture and, according with (%
\ref{coh}), we assume that the exciton field at $t=0$ is described by $%
\langle A_{{\bf 0,}\alpha }\rangle =K_{{\bf 0,}\alpha }(t=0)$ and $\langle
A_{{\bf k\neq 0,}\alpha }\rangle =0$ (all but the ${\bf k=0}$ mode are empty
after the pulse strikes). This approximation is valid for fast pulses and
weak disorder. Since $\widehat{H}+\widehat{U}\,$ does not mix LX and HX, we
solve for each $\alpha $ (${\bf {\it =L,H}}$) the problem of a single
exciton of momentum ${\bf k=0}$ coupled to a continuum of $\alpha $-excitons
at ${\bf k\neq 0}$. The time evolution is given by \cite{Van} 
\begin{eqnarray}
\langle A_{{\bf k\neq 0,}\alpha }(t)\rangle  &=&\frac{\Lambda _{\alpha
}V_{\alpha }({\bf k})}{\delta \Omega _{\alpha }-i\Gamma _{\alpha }}%
e^{-i\omega _{\alpha }t}[e^{-(i\delta \Omega _{\alpha }+\Gamma _{\alpha
})t}-1]  \nonumber \\
\langle A_{{\bf 0,}\alpha }(t)\rangle  &=&\Lambda _{\alpha }e^{-(i\omega
_{\alpha }+i\delta \Omega _{\alpha }+\Gamma _{\alpha })t}  \label{Van}
\end{eqnarray}
where $\Lambda _{\alpha }=\langle A_{{\bf 0,}\alpha }(t=0)\rangle $. $\delta
\Omega _{\alpha }$ and $\Gamma _{\alpha }$ are, respectively, the small
energy renormalization and the decay constant of the state at ${\bf k=0}$
due to $\widehat{U}$. The following conclusions can be drawn from (\ref{Van}%
). First, elastic (disorder-induced) scattering leads to transfer of
coherence from the mode initially excited by the laser to states with ${\bf %
k\neq 0}$. This accounts for the observed light emission in the non phased
matched direction ${\bf k}\neq {\bf 0}$ . Second, within our model, the
scattered field is coherent with the laser field, in agreement with recent
interferometric experiments \cite{recent1}. The intensity of the secondary
emission is given by 
\begin{eqnarray}
&&\langle I\rangle \propto \langle A_{{\bf k\neq 0,}L}^{\dagger }\rangle
\langle A_{{\bf k\neq 0,}L}(t)\rangle +\langle A_{{\bf k\neq 0,}H}^{\dagger
}\rangle \langle A_{{\bf k\neq 0,}H}(t)\rangle +  \nonumber \\
&&\langle A_{{\bf k\neq 0,}L}^{\dagger }\rangle \langle A_{{\bf k\neq 0,}%
H}\rangle +\langle A_{{\bf k\neq 0,}H}^{\dagger }\rangle \langle A_{{\bf %
k\neq 0,}L}\rangle   \label{beat}
\end{eqnarray}
where the last two terms add up to the beating term $\cos [(\omega
_{L}-\omega _{H})t]$ observed in the experiments. It is clear that, within
our model and as for the perfect QW, these Rayleigh beats are due to
interference between the fields associated with the HX and LX polarizations,
which behave as a distribution of {\em classical}{\it \ } antennas. Another
conclusion that can be drawn from (\ref{Van}) is that, at short times, $%
\langle A_{{\bf k\neq 0,}\alpha }\rangle \propto t$. This is in agreement
with the {\em quadratic} ($\propto t^{2}$) rise in the Rayleigh signal
observed at short times and very low exciton densities \cite{Deveaud}, which
is expected when disorder is the only (or the fastest) source of scattering.
The emission for ${\bf k=0}$ decays exponentially with time constant $%
1/\Gamma _{\alpha }$ due to the coupling with the continuum of modes at $%
{\bf k}\neq {\bf 0}$.

As discussed earlier, the perfect QW cannot be described as a collection of
few-level systems because the light creates macroscopic populations of ${\bf %
k=0}$ exciton modes that are uncorrelated. The situation is somehow
different for $\widehat{U}\neq 0$. Here, the wavefunction can be formally
obtained by applying the transformation $A_{{\bf k},\alpha }=\sum_{{\bf \xi }%
}c_{{\bf \xi },\alpha }({\bf k)}B_{{\bf \xi },\alpha }/\sqrt{V}$ so that the
Hamiltonian $\widetilde{H}_{0}=\widehat{H}_{0}+\widehat{U}$ takes on the
diagonal form $\widetilde{H}_{0}=\sum_{{\bf \xi ,}\alpha }\hbar \Omega _{%
{\bf \xi ,}\alpha }B_{{\bf \xi ,}\alpha }^{\dagger }B_{{\bf \xi ,}\alpha }$
( $B_{{\bf \xi ,}\alpha }$ and $B_{{\bf \xi ,}\alpha }^{\dagger }$ are boson
operators and $c_{{\bf \xi },\alpha }({\bf k)}$ are constants). Following
the impulsive excitation, the exciton field is 
\begin{equation}
|\widetilde{\Xi }\rangle \sim e^{-i\widetilde{H}_{0}t/\hbar }\prod_{{\bf \xi 
},\alpha }e^{i\widetilde{K}_{{\bf \xi ,}\alpha }(t)B_{{\bf \xi ,}\alpha
}^{\dagger }}|0\rangle   \label{coh-disorder}
\end{equation}
where $\widetilde{K}_{{\bf \xi },\alpha }=c_{{\bf \xi },\alpha }^{*}({\bf 0)}%
K_{\alpha ,M}/\sqrt{V}$. As (\ref{LH}), $\widetilde{\Xi }$ is a product
state of individual modes that carries a macroscopic {\em polarization} \cite
{SSCus}. Unlike the perfect case, however, the occupation of individual
modes, $\langle B_{{\bf \xi ,}\alpha }^{\dagger }B_{{\bf \xi ,}\alpha
}\rangle $, is not macroscopic since $\widetilde{K}_{{\bf \xi },\alpha }$
does not depend on $V$; see (\ref{K}). Hence, the question arises as to
which extent $\widetilde{\Xi }$ can be distinguished from the sum-state 
\begin{equation}
|\widetilde{\Psi }\rangle \sim \prod_{{\bf \xi }}[1+i\widetilde{K}_{{\bf \xi 
},L}(t)B_{{\bf \xi },L}^{\dagger }+i\widetilde{K}_{{\bf \xi },H}(t)B_{{\bf %
\xi },H}^{\dagger }]|0\rangle   \label{QB-disorder}
\end{equation}
that carries the same polarization and gives the same beats at $\omega
_{L}-\omega _{H}$ (note that $\widetilde{\Psi }$ represents, in some sense, a
collection of atomic 3-level systems with{\em \ randomly-oriented} dipole
moments). The main problem with (\ref{QB-disorder}) is that, since the total
Hamiltonian involves a sum over the light- and heavy-hole modes, the
time-dependent wavefunction must be in a product form. Further, there are
observable differences between states (\ref{QB-disorder}) and
(\ref{coh-disorder}). The two expressions give, in general, different answers
for the mode occupation which is always $<1$ for $\widetilde{\Psi }$ but can
have any value for $\widetilde{\Xi }$ and, moreover, the spectrum of
polarization {\em fluctuations } $\langle P^{2}\rangle -\langle P\rangle ^{2} $
evaluated with $\widetilde{\Xi }$ does not have components at $\omega
_{L}+\omega _{H}$ while $\widetilde{\Psi }$ does. These differences could be
used in the experiments to identify (\ref{coh-disorder}).

In conclusion, light-induced exciton coherence must be understood in terms
of a collective description of the exciton field. This gives product as
opposed to sum states. The resulting LX-HX beats are not due to quantum
mechanical but to classical electromagnetic interference.

We acknowledge discussions with P. Berman, N. H. Bonadeo, B. Deveaud, N.
Garro, S. Haacke, R. Philips, D. Porras, J. Shah, L. J. Sham and D. G.
Steel. Work supported by MEC of Spain under contract PB96-0085, by the
Fundaci\'{o}n Ram\'{o}n Areces and by the U. S. Army Research Office under
Contract Number DAAH04-96-1-0183.

\widetext


\begin{references}
\bibitem{QB}  Leo, K. {\em et al.,} Appl. Phys. Lett. {\bf 57}, 19 (1990);
Feuerbacher, B.F., {\em et al.}, Solid State Commun. {\bf 74}, 1279 (1990);
Smith, G. O. {\em et al.}, Solid State Commun. {\bf 94}, 373 (1995); Ferrio,
K. B. and Steel D. G., Phys. Rev. Lett. {\bf 80}, 786 (1998).

\bibitem{PBvsQB}  Koch, M. {\em et al.}, Phys. Rev. Lett. {\bf 69}, 3631
(1992). Lyssenko, V.G. {\em et al.}, Phys. Rev. B{\bf \ 48}, 5720 (1993);
Zhu, X. {\em et al.}, Phys. Rev. B{\bf 50} 11915 (1994).

\bibitem{Wang}  Wang, H. {\em et al.}, Phys. Rev. Lett. {\bf 74}, 3065
(1995).

\bibitem{Deveaud}  Haacke, S. {\em et al.}, Phys. Rev. Lett. {\bf 78}, 2228
(1997).

\bibitem{Chemla}  Chemla, D.S. , Phys. Today {\bf 46}, 46 (1993).

\bibitem{Stoltzbook}  Stolz, H., {\em Time-Resolved Light Scattering from
Excitons} (Springer, Berlin, 1994).

\bibitem{Baumberg}  Heberle, A.P., Baumberg, J.J. and K\"{o}hler, K., Phys.
Rev. Lett. {\bf 75}, 2598 (1995); Baumberg, J.J. {\em et al.}, Phys. Stat.
Sol (b) {\bf 204}, 9 (1997).

\bibitem{Shah}  Shah, J., {\em Ultrafast Spectroscopy of Semiconductors and
Semiconductors Nanostructures} (Springer, Berlin, 1996).

\bibitem{Cecc}  Guriolo, M.{\em et al.} Phys. Rev. Lett. {\bf 78}, 3205
(1997)

\bibitem{Amand1}  Marie, X. {\em et al.}, Phys. Rev. Lett. {\bf 79}, 3222,
(1997); Le Jeune, P.{\em et al.}, Phys. Stat. Sol. {\bf 164}, 527 (1997).

\bibitem{recent1}  Birkedal, D. and Shah, J. , Phys. Rev. Lett. {\bf 81},
2372 (1998); Woerner, M. and Shah, J. , Phys. Rev. Lett. {\bf 81}, 4208
(1998). Langbein, W., Hvam, J.M. and Zimmermann, R. , Phys. Rev. Lett {\bf 82%
}, 1040 (1999)

\bibitem{recent2}  Kennedy, S.P. {\em et al.}, in Proceedings of ICPS 24,
ed. by Gershoni, D., (World Scientific, Singapore, 1998), to be published;
Garro, N.{\em et al.}, Phys. Rev. B in press.

\bibitem{Glauber}  Glauber, R.J. , Phys. Rev. {\bf 131}, 2766 (1963).

\bibitem{Hopfield}  Hopfield, J.J., Phys. Rev. {\bf 112}, 1555 (1958).

\bibitem{Yajima}  Yajima, T. and Taira, Y. , J. Phys. Soc. Jpn. {\bf 47},
1620 (1979); Haroche, S., in {\em High Resolution Laser Spectroscopy}, ed.
by Shimoda, K. (Springer, Berlin 1976), p. 253.

\bibitem{QD}  Bonadeo, N. H.{\em et al.}, Science {\bf 282} 1473 (1998).

\bibitem{SR}  Haug, H. and Schmitt-Rink, S., Prog. Quant. Electr. {\bf 9}, 3
(1984); Wegener, M. {\em et al.}, Phys. Rev. A{\bf 42}, 5675 (1990).

\bibitem{Comte}  Comte, C. and Mahler, G. Phys. Rev. B {\bf 34}, 7164 (1986).

\bibitem{Jorda}  See,{\em \ e.g.}, Jorda, S. , Rossler, U. and Broido, D.,
Phys. Rev. B {\bf 48}, 1669 (1993), and references therein.

\bibitem{Galindo}  Galindo, A. and Pascual, P., {\em Quantum Mechanics}
(Springer, Berlin, 1991).

\bibitem{JOCM}  Fernandez-Rossier, J., Tejedor, C. and Merlin, R., J.
Phys.Cond. Matt, in press.

\bibitem{phonon}  Hase, M.{\em et al.}, Jpn. J. Appl. Phys. {\bf 37}, L281
(1998).

\bibitem{teoray}  Zimmermann, R., Nuovo Cimento D {\bf 17},1801 (1995).

\bibitem{Van}  Van Kampen, P., {\em Stochastic Processes in Physics and
Chemistry} (North Holland, Amsterdam, 1992).

\bibitem{SSCus}  Fernandez-Rossier J., Tejedor, C. and Merlin R., Solid
State Commun. {\bf 108}, 473 (1998).
\end{references}
\end{document}